\documentstyle[aaspp4,epsf]{article} 

\def\edcomment#1{\iffalse\marginpar{\raggedright\sl#1\/}\else\relax\fi}
\reversemarginpar

\begin{document}
\title{Scattered Light from Close-in Extrasolar Planets: \\
Prospects of Detection with the MOST Satellite}

\author{Daniel Green, Jaymie Matthews}
\affil{Department of Physics and Astronomy, University of British Columbia,
Vancouver, BC, Canada, V6T 1Z1}
\author{Sara Seager}
\affil{Department of Terrestrial Magnetism, Carnegie Institution of
Washington,
5241 Broad Branch Rd. NW, Washington, DC 20015, USA}
\author{Rainer Kuschnig}
\affil{Department of Physics and Astronomy, University of British Columbia,
Vancouver, BC, Canada, V6T 1Z1}

\begin{abstract}
The ultra-precise photometric space satellite MOST (Microvariability
and Oscillations of STars) will provide the first opportunity to
measure the albedos and scattered light curves from known short-period
extrasolar planets. Due to the changing phases of an extrasolar planet
as it orbits its parent star, the combined light of the planet-star
system will vary on the order of tens of micromagnitudes.  The
amplitude and shape of the resulting light curve is sensitive to the
planet's radius and orbital inclination, as well as the composition
and size distribution of the scattering particles in the planet's
atmosphere.

To predict the capabilities of MOST and other planned space missions,
we have constructed a series of models of such light curves, improving
upon earlier work by incorporating more realistic details such as:
limb darkening of the star, intrinsic granulation noise in the star
itself, tidal distortion and back-heating, higher angular resolution
of the light scattering from the planet, and exploration of the
significance of the angular size of the star as seen from the planet.
We use photometric performance simulations of the MOST satellite, with
the light curve models as inputs, for one of the mission's primary
targets, $\tau$ Bo\"otis.  These simulations demonstrate that, even
adopting a very conservative signal detection limit of 4.2 $\mu$mag in
amplitude (not power), we will be able to either detect the $\tau$
Bo\"otis planet light curve or put severe constraints on possible
extrasolar planet atmospheric models.
\end{abstract}

\section{Introduction}

Since the discovery of a planet around 51 Peg b in 1995 (Mayor \&
Queloz 1995), the field of extrasolar planetary research has grown
steadily.  Radial velocity surveys (e.g., Marcy et al. 2003; Santos et
al. 2000; Tinney et al. 2002) have found over 100 extrasolar giant
planets (EGPs) orbiting nearby stars. In addition, $\sim$ 10 of these
systems contain planets with semi-major axes $\lesssim 0.05$~AU, here
after called close-in EGPs (CEGPs).  The radial velocity surveys
provide the planet's minimum mass and orbital parameters (such as
semi-major axis and eccentricity) but nothing else about the planet's
properties.  Despite the growing numbers of discoveries, we only know
detailed and accurate properties of a single extrasolar planet: HD
209458b. Observations of this transiting planet HD 209458b (Charbonneau
et al. 2000; Henry et al. 2000; Brown et al. 2001) have provided
measurements of the radius and mean density of the planet, providing
the first information on the planet's composition. Furthermore, the
detection of the trace element sodium in HD209458b's atmosphere by
Charbonneau et al. (2002) has provided the first constraint on an
extrasolar planet's atmosphere.

Ongoing planet transit searches (see Horne 2003) should increase the
number of extrasolar planets with observed physical properties by
providing a measured radius and inclination for discovered
planets---see Konacki et al. (2003) for a description of the first
planet detected with the transit search method.  However, even for
transits, little information is coming directly from the planet.  For
non-transiting planets, direct spectroscopy and photometry appear to
be the most likely sources of additional information.  Spectroscopy
could reveal atmospheric composition. Ultra-precise photometry has the
potential to reveal the nature of the atmospheric scattering
particles: due to the changing phases of a short-period extrasolar
planet as it orbits its parent star, the combined light of the
planet-star system will vary on the order of tens of micromagnitudes.
The shape of the resulting light curve is indicative of the
atmospheric scattering particles' composition and size distribution.
Unfortunately, ground-based photometry is limited by atmospheric
scintillation to detect magnitude variations of $10^{-4}$.  Such
precision is generally only possible for bright variable stars with
periods of only a few minutes, such as the rapidly oscillating Ap
(roAp) stars, where long-term drifts do not interfere with signal
detection at high frequencies.  An example of the state of the art in
rapid ground-based photometry is the work of Kurtz et al.  (2003), who
set a threshold of 0.2 millimag in their null detection of
oscillations in the Ap star HD 965.  For periods of days, we know of
no photometric measurements that have achieved this level of
precision. However, assuming a grey albedo, an upper limit of 
$5 \cdot 10^{-5}$ for the variation of the flux ratio 
has been establish for the $\tau$ Bo\"otis system
(Charbonneau et al. 1999, Leigh et al. 2003).  
The planetary light curve amplitudes are anticipated to be
below this threshold.  Nevertheless, planned space-based
photometric telescopes are expected to detect $\mu$mag variations in
the next few years.

The first of these missions to go into orbit should be MOST
(Microvariability and Oscillations of STars)---a Canadian Space Agency
microsatellite housing an ultra-precise photometric instrument, to be
launched on 30 June 2003.  MOST was designed to detect and
characterize rapid acoustic oscillations in solar-type stars, but it
also has the potential to measure the scattered light from known
CEGPs.  The light curve data are sensitive to a planet's radius,
inclination, and most importantly albedo, which in turn depends on the
thermal equilibrium of the planet and the composition and size range
of the primary scattering particles.

MOST will not be searching stars for new planets, due to its small
aperture and limited number of accessible targets, but rather
monitoring stars already known to have CEGPs, searching for scattered
light signals whose periods are already well determined.  The goal is
to detect the scattered light signature of an extrasolar planet for
the first time, and to provide empirical data to test models of CEGP
atmospheric composition.  The MOST target list includes three stars
with CEGPs in its first two years: 51 Peg, $\tau$ Bo\"otis, and
HD~209458.  The photometric data will also be used to search for
solar-type oscillations in the parent stars, whose eigenspectra can
better refine their masses and main-sequence ages.  This will also be
extremely valuable in understanding the nature and history of the
CEGPs themselves.

Other funded space missions---COROT (CNES/ESA 2005), Kepler (NASA
2007) and Eddington (ESA 2007)---will monitor fields of tens of
thousands of stars, discovering hundreds of new EGPs by their
scattered light curves (in addition to their primary extrasolar planet
goal of searching for transiting Earth-sized planets).  MOST will
provide a valuable starting point for these missions by determining
the signature of the CEGP light curves that can then be used for
detection algorithms and also by characterizing the low-amplitude
photometric variability of solar-type stars, which will affect planet
light curve and transit detections. A recent paper by Jenkins \& Doyle
(2003) evaluates Kepler's ability to discover CEGPs by their light
curves, around stars without known planets. Their paper includes an
estimate of the number of planets Kepler expects to detect and a
description of detection algorithms. Our paper is complementary,
describing MOST's potential for detecting CEGP light curves
of {\it known planets} with known orbital periods.

Using a Monte Carlo method, Seager, Whitney, \& Sasselov (2000) first
generated scattered light curves for generic close-in EGPs to show
that the resulting light curves were highly dependent on the
composition and size distribution of the condensates in the
atmosphere.  Furthermore, Seager et al. (2000) showed that systems
like 51 Peg might show light variations as large as 60 $\mu$mag
peak-to-peak.  Even signals twenty times smaller are expected to be
within the range of detectability by MOST and other space missions, so
these early results inspired the MOST team to expand their science
mission to include CEGPs.

In this paper, we present the results of physically more complete
models of CEGP scattered light curves that include various types of
noise, and we evaluate MOST's capability to detect them.  In \S2, we
describe the planet atmosphere model and the Monte Carlo model used to
produce the synthetic planet light curve data.  The stellar noise
model is described in \S3 and the MOST performance simulation in \S4.
In \S5 we present preliminary results and discussion of both the model
and the simulated MOST data.  We conclude the paper with a discussion
of future prospects in \S6.

\section{The Planet Scattered Light Curve Model}
\label{sec-atm}
\subsection{The Atmosphere Structure Model}

The 3D Monte Carlo (MC) model aims to compute the emergent flux at
visible wavelengths from starlight that has anisotropically scattered
through the planetary atmosphere. In order to compute the photons'
paths through the atmosphere an input atmospheric structure is
needed. For the MC code purposes, this input atmospheric structure
consists simply of the wavelength-dependent absorption and scattering
coefficients as a function of location in the atmosphere.  For
simplicity we consider a homogeneous atmosphere in which case only a
1D radial profile (i.e., as a function of vertical atmospheric depth)
of absorption and scattering coefficients is needed.  (The 3D MC code
is required because of highly anisotropic scattering properties of
some condensate particles.)  Computing the radial distribution and
abundance of all of the different absorption and scattering
coefficients themselves is a complex task and depends on temperature,
pressure and chemical abundances, as described below.

\subsubsection{Description of The Model Atmosphere}

The atmosphere model used here is a 1D plane-parallel radiative +
convective equilibrium code. Full details are described in Seager
(1999), Seager et al. (2000), and Seager \& Sasselov (in preparation).
Three parameters that describe the atmosphere are solved from three
equations in a Newton-Raphson type scheme. The three parameters are
temperature (as a function of depth), pressure (as a function of
depth) and the radiation field (as a function of wavelength and
depth).  The three equations are radiative transfer, radiative +
convective equilibrium, and hydrostatic equilibrium.  These equations
and the three parameters are highly coupled, which is why to compute
the temperature-pressure structure we must also simultaneously solve
for the radiation field.  In order to solve the three atmosphere
equations, an upper and lower boundary condition are needed. The upper
boundary condition is the flux from the parent star computed with
Kurucz model atmospheres (Kurucz 1992) and the lower boundary
condition is the flux from the interior of the planet. We assume that
the heating from irradiation is instantaneously redistributed around
the tidally-locked planet (but cf. \S\ref{sec-uncertainties}).

Beyond the equations and boundary conditions there are several other
inputs to the model atmosphere code.  These include planet semi-major
axis and surface gravity. We adopt solar abundances to provide an
easily comparable standard for future models.  The choice of abundance
value is a smaller uncertainty compared to cloud opacity (see
uncertainties described in \S\ref{sec-uncertainties}).  In addition we
do not know if the origin of Jupiter's high metallicity also applies
to extrasolar planets; therefore a higher than solar metallicity is
not a suitable reference point. The number density of gas and solid
species comes from a Gibbs free energy chemical equilibrium
calculation (described in Seager et al. (2000)) which specifies the
species abundance as a function of temperature and pressure.  The
opacities for H$_2$O, CH$_4$, Na, K, and pressure-induced H$_2$-H$_2$
and H$_2$-He and MgSiO$_3$ are used.  Note that H$_2$O is the most
important gas in determining the temperature-pressure structure due to
stellar irradiation. Full references for the opacities used are listed
in Seager et al. (2000). Note that our more recent work (Seager \&
Sasselov in preparation) shows that use of more recent opacities
results a similar temperature-pressure structure to the one used here
(Seager \& Sasselov, in preparation), certainly similar enough for the
goal of this paper. With a self-consistent solution for the vertical
temperature-pressure profile in a plane-parallel atmosphere, the
absorption and scattering coefficients used in the Monte Carlo
calculations come directly out of the calculation.

\subsubsection{Model Uncertainties}
\label{sec-uncertainties}
Our model is relevant to first order and more than sufficient for this
paper's primary goal of computing signatures of extrasolar planets with
real instrumental and stellar noise concerns. Nevertheless we must keep in
mind that there are many uncertainties in the model and any specific
model can involve many choices for input parameters. Ultimately
the MOST data will be able to constrain the large choice of parameter
space and help narrow down the uncertainties.

Recent calculations of atmospheric circulation (Guillot \& Showman
2002; Showman \& Guillot 2002; Cho et al. 2003) have shown that the
stellar irradiation acting on a close-in tidally-locked gas giant
planet could cause a highly non-uniform temperature distribution with
horizontal temperature variations of up to 1000~K.  While such
atmospheric circulation models are not yet sophisticated enough to
generate temperature-pressure profiles and emergent spectra they do
indicate a major uncertainty of all current CEGP atmospheric structure
models that needs to be addressed in the near future.  Even though the
scattered light curves depend on the illuminated side only, the
atmospheric circulation is still necessary to compute the atmospheric
structure consistently.

There are many other uncertainties in the atmospheric
models. High-temperature condensates such as MgSiO$_3$, Al$_2$O$_3$,
and Fe are stable at the expected CEGP atmosphere temperatures and
pressures (see e.g., Seager et al. 2000). These high-temperature
condensates form clouds, just as water ice does here on Earth. These
clouds present a major complication for EGP modeling because the
strong condensate opacity is highly sensitive to the composition and
size distribution of particles.  The size distribution is determined
by a number of physical processes that compete for grain growth and
grain destruction, including condensation, coalescence, sublimation,
and sedimentation. Two recently developed cloud models (Ackerman \&
Marley 2001; Cooper et al. 2003) aim to predict particle sizes and are
meant to be used consistently in a model atmosphere that determines
the temperature, pressure, and radiation field (e.g., Marley et
al. 2002).  Nevertheless even these cloud models are used in
homogeneous layers, not patchy clouds that may exist, and the models
still have other uncertainties.  We use the results of such
computations as a basic guide for our choice of cloud particle size.
Other uncertainties are about upper atmosphere processes such as
photoionization and photochemistry which could cause small absorptive
particles.

\subsubsection{The Fiducial Atmosphere Models}

We must make choices within the large model atmosphere input parameter
space; here we have chosen to work with two fiducial models. Both
models have solar abundance.  The first model is our cloudy model
where we choose a vertically and horizontally homogeneous cloud of
MgSiO$_3$ cloud particles that is two pressure-scale heights
thick. Note that even though the cloud parameters are hard-wired, the
cloud's vertical location is self consistently solved for according to
the temperature-pressure saturated vapour pressure relation.  The
condensates are prescribed to have a log-normal particle size
distribution having mean radii of 5~$\mu$m and
$\sigma=1.5$~$\mu$m. The phase function (i.e., the directional
scattering probability) of the condensates is computed with a Mie
scattering code (see Figure 1). This silicate cloud model is motivated
by considering chemistry models (see Fegley \& Lodders 1996) that show
MgSiO$_3$ is likely to form first as the planet cools at the expense
of other Mg species.  In addition, MgSiO$_3$ is likely to be the
``top'' cloud that the stellar photons will reach first. This is
because MgSiO$_3$ is likely to be the lowest-temperature condensate
at the relevant CEGP temperatures.

For comparison we use a second fiducial model of a cloud-free planet
where the scattering is due to Rayleigh scattering mostly from gaseous
H$_2$.  The case of no condensates on the dayside may be realized in
some cases of atmospheric circulation where condensates are
transported to a much cooler night side where they settle out
permanently from the atmosphere (Guillot \& Showman 2002). For more
details of the temperature-pressure profiles and the corresponding
emergent spectra see Seager \& Sasselov (in preparation).

\subsection{Monte Carlo Model}

Our Monte Carlo model is based on the methods presented in Code \&
Whitney (1995) and Seager et al. (2000). The overall Monte Carlo
scattering problem involves following photons that come from a
star, enter the planet atmosphere at a given location traveling
into a given direction, scatter repeatedly in the planetary
atmosphere, and finally exit the planet atmosphere.
Essentially, probability distributions are produced for all
factors involved in the photon scattering problem (e.g., initial
position, distance between interactions, absorption vs.
scattering) and are sampled according to
\begin{equation}
\xi =\int_{0}^{a} p(x) dx,
\end{equation}
where $\xi$ is a random number between 0 and 1, $p(x)$ is the
probability density and $a$ is the output value.
The process is repeated for each photon individually. Over 50 million
photons were used in each run in order to ensure low statistical
error.  The final photon counts are normalized to give a ratio of
the reflected flux from the planet to the flux from the star (flux ratio).
One attractive feature of this method is that since each photon is
independent of the last, the algorithm running time is linearly
dependent on the number of photons used. Although our code lacks
efficient algorithms it is still capable of simulating large numbers
of photons in relatively short periods of time.

\subsubsection{Initial Photon Properties}

Before a photon is scattered for the first time initial
characteristics of the photon are determined. {\it The photon
wavelength} is not calculated exactly, but instead a random number
determines which range of wavelengths it falls into according to
the blackbody spectrum of the star. The MOST bandpass is close to
a box function from 400 nm to 750 nm and we have chosen ten
wavelength bins to represent this bandpass (based on the planet
atmospheric spectrum generated in \S 2.1).

{\it The initial coordinates and trajectory} of the photon must
also be generated.  The starting coordinates are produced by
generating random $x$, $y$ coordinates on a disk of the same radius as
the planet. The EGPs with measurable scattered light curves will have
semi-major axes within 0.1 AU; the stellar flux is not
plane-parallel and so the initial trajectory of the photon is
non-trivial to determine. The initial trajectory of the photon is
determined from the probability distribution
\begin{equation}
1=k \int_{0}^{R_s} {2\over 5} \pi r \left(2+3 \cos \left[\arcsin {r\over
R_s}\right]\right) dr,
\end{equation}
which includes an approximation of solar limb darkening (Carroll \&
Ostlie 1996). In this equation, $r$ is the radial
distance on a disk of radius $R_s$ (where $R_s$ is the parent star
radius), and $k$ is the normalization constant. This distribution only
approximates the relative amount of light incident from different
directions.

\subsubsection{Photon Scattering, Absorption, and Flux}

Once the photon enters the atmosphere, it is followed through all
scattering processes until it exits the atmosphere or is absorbed by a
gas or solid particle. Distances between interactions are calculated
separately for all types of events (scattering by gas, scattering by
solids, absorption by gas, absorption by solids), and the shortest
distance sampled produces an interaction. For scattering events, the
new trajectory is determined by sampling the appropriate phase
function (see Figure 1 for the phase function of MgSiO$_3$).

Following these interactions, the same distance calculations are
repeated until the photon is absorbed or exits the atmosphere. If
absorption occurs, we assume the photon vanishes from the MOST
bandpass; absorbed photons will be reemitted at IR wavelengths where
the CEGP thermal flux peaks. Although the absorbed photons do affect
the heat balance of the planet, this is already taken into account
from our atmospheric structure models described in \S2.1.1--2.1.3. If
the photon escapes the atmosphere, it is binned according to
trajectory angle relative to the direction of the parent star
($z$-axis).  This binning method assumes symmetry about the $z$-axis;
reasonable for a symmetric atmosphere with symmetric illumination.

After the pre-specified number of photons have been sent through the
atmosphere, the photon counts are normalized to give the emergent
flux.  At intervals given by expected integration times for MOST of
1-2.5 minutes, the orbital position is calculated from the
eccentricity, inclination, semi-major axis and orientation of the
orbit. From the orbital position the phase angle (the
star-planet-observer angle; in our case the angle between the $z$-axis
and the observer) is calculated. The flux for the phase angle is taken
from the binned data and normalized for the current distance.

\subsubsection{Planet Tidal Distortion Effects on the Light Curve}
Sinusoidal modulations in photometric light curves are observed in
binary star systems with tight orbits due to tidal distortion of the
stars into ellipsoidal shapes (Von Zeipel 1924; Kitamura \& Nakamura
1988).  Even short-period (2-day) companions to solar-type primaries
can cause a gravitational distortion visible on millimag light curves
if the companion mass is at least $\sim$ 0.2 $M_{\odot}$ (Drake 2003).
We examine the same effects from a tidally distorted planet, to see if
they will affect the scattered light curve at the micromagnitude
level, by estimating the distorted length of the planet's axes under
the assumption of an isothermal expansion.  Assuming equilibrium and
cylindrical symmetry,
\begin{equation}
R_{z} = {{\left[\ln\left({{g-2 g_{t}}\over {g}}\right)-{{m_{H}g R_{p}}\over
{k_{b} T}}\right]  {k_{b} T}}\over {m_{H} (g-2 g_{t})}},
\end{equation}
and
\begin{equation}
R_{x} = {{\left[\ln\left({{g+ g_{t}}\over {g}}\right)-{{m_{H}g R_{p}}\over
{k_{b} T}}\right] {k_{b} T}}\over {m_{H} (g+ g_{t})}},
\end{equation}
where $R_{z}$ is the radius of the planet in the direction of the
star, $R_{x}$ is the perpendicular radius and $R_{p}$ is the initial
spherical size of the planet.  Furthermore, $g$ is the surface
gravity, $m_{H}$ is the mass of hydrogen and $ k_{b} T$ is the thermal
energy. The tidal acceleration is $g_{t} = {G M_{s} R_{p}\over
{a^{3}}}$, where $M_{s}$ is the stellar mass, $a$ is the semi-major
axis and $G$ is the gravitational constant.  Given the phase angle,
the relative increase or decrease in intensity is added through
geometric optics instead calculating it directly in the Monte Carlo
code.  This approach is approximate, but given the relatively small
size alterations the impact of this assumption should be minimal. From
the above equations, we found the tidal distortion to be a $\sim$
10$^{-7}$ effect, well below the $\sim$ 10$^{-5}$ light curve signal.

\subsubsection{Back Heating}
CEGPs are expected to be tidally locked due to tidal interactions
between the planet and the star (Goldreich \& Soter 1966; Guillot
et al. 1996). The stellar atmosphere may also be affected by tidal
interactions with the CEGP. Combined with the back scattering of
light from the planet, the stellar atmospheres could potentially
have a flux hot spot with a rotational variation of the same
period as the planet's rotational period. Although stellar back heating is
expected to be a small effect, it could be important because it
would share the same period as the planetary light curve.

To investigate the magnitude of the back heating effect we
construct an approximate model, based on the results of the Monte
Carlo code. From the distance between the planet and the star (it
is assumed that only roughly circular orbits will produce the back
heating effect) and the angular binning used in the scattering
code, several rings on the stellar surface are considered. Each
stellar ring is considered to initially radiate energy according
to the black body equation $E = A\sigma T_{e}^{4}$, where $A$
is the emissivity, $\sigma$ is the Stefan-Boltzmann constant and
$T_{e}$ is the effective temperature. Using the
same equation for the total flux emitted by the star, an estimate
of the scattered energy is produced from the fraction of the light
scattered from the planet back towards the star.  Since the star
must reach equilibrium between the input and expelled energy at
each point on the surface, a ratio of the intensity of flux on
each ring to the average can be made.  Making the approximation
that the flux at the planet is unchanged, we get
\begin{equation}
{F_{R} (\theta)\over F_{AVG}} =
1+{P(\theta)\over Z} \pi R_{s}^{2} {R_{p}^{2}\over a^{2}} {1\over A_{R}},
\end{equation}
where $R_{s}$, $R_{p}$ are the star and planet
radius, $a$ is the planet semi-major axis, $Z$ is the total number of
photons in the Monte Carlo code, while $P(\theta)$ is the number
of photons reflecting into angle $\theta$. $A_{R}$ is the area of
the ring of the stellar surface produced by binning at $\theta$,
$F_{AVG}$ is the initial flux of the ring and $F_{R}$ is the
increased flux due to scattered light.  This ratio is calculated
for each ring on the stellar surface.

Once the increased stellar flux is calculated, a light curve is
produced for each ring using the limb darkening profile of the star.
The profile is constructed using an approximation for solar limb
darkening (Carroll \& Ostlie 1996). This limb darkening approximation
has good agreement with measurements averaged over the visible
spectrum. The increase in stellar flux is spatially integrated over
the stellar disk for each time interval. Because of the small size of
the scattered light ratio and $R_{p}^{2}\over a^{2}$, our estimate
shows back heating to be a small effect (on the order of 10$^{-7}$)
compared to the $\sim 10^{-5}$ CEGP light curve.

\section{Stellar Granulation Noise}

One of the fundamental limiting factors in the spectroscopic detection
of extrasolar planets through Doppler shifts is the intrinsic radial
velocity
noise due to the changing pattern of rising granules at the top of the
convection zone.  The variation in filling factor and contrast of the
granulation pattern is also an important noise source in ultra-precise
photometry of solar-type stars.

The level of granulation noise is correlated with chromospheric
activity, which in turn depends on stellar rotation rate, surface
magnetic activity, as well as depth of the surface convection
zone. The sample, selected for Doppler searches for extrasolar planets tend
to
be chromospherically quiescent, so the targets for MOST photometry
will also share that trait.  However, granulation noise may still be
the dominant noise source, especially at low frequencies.

Granulation noise is non-white, and photometry of the Sun suggests
that the noise spectrum has an approximate $1/f$ dependence of
amplitude on frequency (e.g., Kjeldsen \& Frandsen 1992; Kjeldsen \&
Bedding 1998).

To simulate this noise source, we generate a grid of frequencies from
zero to the Nyquist frequency appropriate to the simulated data
sample.  These values are inverted to create an array of $1/f$ values,
then multiplied by a corresponding array of random numbers
(distributed normally about zero with a variance of one) to randomize
the amplitudes and phases of the components of the intrinsic noise.
An inverse discrete Fourier transform on the resulting array yields a
synthetic time series of granulation noise.  This time series can then
be multiplied by a scaling factor to match the overall level of
granulation noise to be introduced.

For the Sun, photometric granulation noise at a frequency of 0.1 mHz
is approximately 2 parts per million in integrated optical broadband
light (see, e.g., Kjeldsen \& Bedding 1998). We have been guided by
this in our simulations, since ground based photometry of other
solar-type stars does not set useful upper limits on the granulation
noise at relevant frequencies.

\section{MOST as an Ultraprecise Photometer}

MOST is a small optical telescope (aperture = 15 cm; Maksutov design),
with a single broadband filter ($350 \leq \lambda \leq 700$ nm),
feeding a CCD photometer, aboard a microsatellite platform (mass = 54
kg; dimensions $60 \times 60 \times 25$ cm). The microsat will be
stabilized to a pointing accuracy of about $\pm10$ arcsec by a set of
miniature low-power reaction wheels designed and built by Dynacon
Enterprises Ltd. of Toronto, Canada.  Although this level of attitude
control outperforms (by a factor of several hundred) any existing
microsat with such small inertia, it is still relatively poor pointing
for an astronomical instrument. Hence, the MOST photometer is equipped
with an array of Fabry microlenses to project fixed images of the
entrance pupil of the telescope, illuminated by the target starlight,
onto the Science CCD.  Unlike a wandering image of the star, this
extended Fabry image (covering about 1400 pixels) of the CCD makes the
collected signal quite insensitive to the flatfield sensitivity
gradients of the detector, even at the sub-pixel scale. For more
details about technical aspects of the MOST experiment, see Walker,
Matthews et al. (2003).

MOST will be launched into a low-altitude (820 km) circular polar
orbit, whose slight inclination will cause it to precess at the
sidereal rate, so the orbital plane is synchronous with the Sun.
Launch is scheduled for 30 June 2003 aboard a Russian three-stage
``Rockot'' launch vehicle (designated an SS-19 in the West, since it
is a former Soviet ICBM) from the Plesetsk Cosmodrome.  It will be
injected into an orbit which will keep it above the Earth's
terminator. From this vantage point, the telescope will always look
over the shadowed limb of the Earth, minimizing scattered Earthlight
which could interfere with the ultraprecise photometry.  This orbit
also provides a Continuous Viewing Zone (CVZ) in the sky spanning
declinations $+34^{\circ} \leq \delta \leq -18^{\circ}$.  Stars
passing through the centre of this band will remain visible
continuously for about 8 weeks.  The MOST CVZ includes several
prominent extrasolar planet systems, including 51 Pegasi, $\tau$
Bo\"otis, and HD209458, which have been included as primary science
targets for the MOST mission.

MOST was designed to achieve the mission's primary goal of detecting
rapid photometric oscillations (periods of several minutes) in bright
($V \leq 6$) solar-type, metal-poor subdwarf and magnetic Ap stars
with precisions approaching 1 part per million (1 $\mu$mag).  Although
MOST is a {\em non}-differential photometer, the relatively high
frequencies of the periodic oscillations can be clearly distinguished
in a Fourier spectrum of the data from the lower-frequency
modulations, drifts and noise (e.g., orbital variations with $P_{\rm
orb} \simeq 100$ min; granulation noise in the stars themselves).

This is not true for the periodic reflected light signals from
extrasolar planets, whose orbital frequencies ($\nu_{\rm orbit} \simeq
0.2 - 0.3$ d$^{-1}$ $\simeq 0.003$ mHz) are very low compared to the
intended sensitivity range of MOST ($\nu_{\rm osc} \simeq 0.5 - 6$
mHz).  Therefore, MOST non-differential photometry of extrasolar
planet systems will be more prone to the long-term drifts and
modulations. If MOST were intended to be a planet {\em hunter},
searching this low-frequency regime for unknown periodic signals in a
noisy background, this might be a fatal flaw. However, as a probe of
known extrasolar planet systems whose periods have already been
specified accurately from radial velocity data, MOST can be quite
effective, as we will demonstrate in \S5.

\subsection{Modeling the Photometric Performance of MOST}

MOST is optimized to collect very precise photometry for stars in the
magnitude range $0.0 \leq V \leq 6.0$, with integration times from
about 0.2 sec to 60 sec depending on the flux of the target.  For
extrasolar planet photometry, the integration time would be set to
bring the total signal per exposure to about 80\% of the full-well
potential of each CCD pixel, maximizing S/N without sacrificing
linearity. For a star of magnitude $V = 0$, MOST would collect
approximately $1.6 \times 10{^8}$ electrons/sec, although to avoid
saturation on such a bright target, the integration time would have to
be about 0.2 sec. For extrasolar planet targets with long periods
(compared to the rapid stellar oscillations), fast time sampling is
not a consideration so every exposure can be long enough to guarantee
a maximum S/N of about 220 per pixel per exposure; hence, a S/N of
about 8300 over the entire 1400-pixel Fabry image.  Further
improvements in S/N are possible by substantially binning these short
exposures. With extrasolar planet orbital periods of several days,
60-sec exposures can safely be binned in groups of several hundred
without appreciably losing resolution in orbital phase.

As part of the design and testing process for the MOST mission, a
comprehensive simulator of MOST photometry was developed, written in
IDL (Kuschnig et al. 2003).  This simulation code was designed to
include as many noise, drift and modulation effects as could be
anticipated and modeled by the MOST Instrument and Science Teams.  The
effects can be grouped into four categories: (1) intrinsic variations
and noise from the target star (and planet); (2) orbit and radiation
environment; (3) sky background and attitude control errors; and (4)
detector and electronics.  These will be discussed in more detail by
Kuschnig et al. (2003) but are summarized in the next few paragraphs.

{\bf 1. Effects intrinsic to the target star + planet.} These include
the Poisson noise associated with the total flux from the system,
photometric noise associated with granulation in the star's
photosphere, rotational modulation due to starspots, and the periodic
variations in scattered light from the planet. The last three are
included in the extrasolar planet light curve model (see \S2),
although they can also be introduced by the MOST photometry simulator
independently.

{\bf 2. Orbit and radiation environment.} MOST will circle the Earth
approximately every 100 minutes; the exact period will be known very
accurately after launch and final orbital injection.  Although from
its vantage point above the terminator, MOST will only see the
nightside limb of the Earth in normal operation, it is possible that
there will be some contamination due to stray light scattered from the
Earth, varying with MOST's orbital period. It is possible to add stray
light at a level consistent with albedo models of the Earth (e.g.,
Shaw et al. 1998; Buzasi 2002).  The fluxes of high-energy protons and
electrons have been calculated for the MOST orbit, and cosmic ray hits
onto the Science CCD based on these fluxes have been included in the
simulations. Also, MOST will pass through the South Atlantic Anomaly
(SAA), exposing it to much higher particle fluxes for several minutes
on some orbits. Rather than try to extract photometry from the CCD
during these brief passages, we have conservatively not included these
data in the time series, introducing short non-periodic gaps which
have only a modest effect on the window function of the Fourier
spectrum.  The CCDs are temperature stabilized by a passive thermal
control system which maintains the operating temperature at about
$-40.0 \pm 0.1$ C. However, we have anticipated there might be a
subtle modulation in temperature of 0.1 C (the level of control of the
CCD thermal control system) at the MOST orbit period, and have
included that effect on the CCD output.

{\bf 3. Sky background and attitude control errors.} The MOST
photometry is obtained through a diaphragm 1 arcminute in diameter,
which will include a sky background of Zodiacal light, atomic oxygen
glow (even at 820 km altitude), stray Earthlight (already modeled in
category (2)), scattered light from off-axis sources, and faint stars
and galaxies adjacent to the target in the sky. The MOST Telescope and
Camera are equipped with a series of baffles designed to reduce
parasitic stray light by a factor of $10^{-12}$, but we conservatively
include a variable sky background.  Wander in telescope pointing due
to attitude control errors of about $\pm10$ arcsec has several
effects: (a) The target starlight beam wanders across the surface of
the Fabry lens which produces the pupil image, subtly changing the ray
paths within the glass and possibly encountering contaminants on
different parts of the lens surface. (b) Faint stars or galaxies near
the edge of the diaphragm can wander in and out of the field, varying
the sky background level. (c) The pupil image will not be completely
fixed, although the image motion will only be at a level of about 0.1
pixel in the MOST focal plane.  The attitude control system (ACS)
errors are modeled based on simulations of the satellite pointing
performance. These models are used to introduce errors due to the
target starbeam wander (effect (a)). Effect (b) is negligible for the
bright targets we consider here, unless a background star is within
about 8 magnitudes of the target star brightness.  We have
investigated all the target fields -- including the extrasolar planet
fields -- and there are no potentially worrisome neighbours in any
field. Effect (c) is negligible because of the large size of the pupil
image on the CCD, so even sub-pixel sensitivity variations of a few
$\times 10\%$ do not manifest themselves in the pupil image motion.

{\bf 4. Detector and electronics.} These effects include: (a) CCD
readout noise; (b) dark noise and possible drifts in dark current; (c)
pixel-to-pixel and sub-pixel sensitivity gradients (see (3) above);
(d) analogue-to-digital conversion (ADC) non-linearities; (e) slight
variations in readout-channel gain; and (f) uncertainties in the
integration times.

In our simulations, by far the dominant sources of noise are Poisson
statistics (photon noise) and stellar granulation.  The photon noise
in the $\tau$ Bo\"otis simulations shown in \S5 is at a level of 0.74
ppm (1$\sigma$). If granulation noise is included in the simulations,
the 1-$\sigma$ noise rises to 1.40 ppm.  The other noise sources
turn out to be negligible for the timescales associated with CEGP
scattered light curves.  However, for a more detailed breakdown of
the photometric error budget of MOST, see Tables 5 and 6 in Walker,
Matthews et al. (2003).

\section{Results}

\subsection{Simulations of Scattered Light Curves}

One of the most important aspects of our model is taking into account
the angular size of the star as seen from the planet.  We have found
that using an extended source with limb darkening, the shape of the
resulting light curve is significantly altered (see Figure 2) compared
to a point source.  With a point source, the detailed, high angular
resolution features of the phase function of scatterers remain
apparent at planet-star separations $\gtrsim$ 0.07 AU.  This effect
had gone unnoticed in previous simulations (Seager et al. 2000)
because of the low angular resolution used to calculate the fluxes
scattered from the planet in those models.  Our work has shown it is
essential to use angular bins of less than a degree to properly
compute the light curves, especially for the very close-in
extrasolar planets. With the proper source geometry, these features
are smoothed for separations smaller than 0.07 AU and the amplitude of
the light curve is reduced by up to 20\% for orbital inclinations near
$90^{\circ}$ (as noted in Seager et al. 2000).

The addition of stellar back heating was found to be negligible even
for space-based photometry of precision 1 ppm. Typically, stellar back
heating contributed a flux ratio of $10^{-7}$ or less.  The effect of
tidal distortion was slightly larger than back heating.  The tidal
distortion alone can change the scattered flux of a planet at 0.045 AU
from 0.9956 at minimum projected area to 1.0022 at maximum area, where
1.0 is the undistorted value.  This translates to an additional
variation of $5 \times 10^{-7}$ about the mean in the light curve.
Although treatment of both these effects was approximate, our initial
estimates suggest that their influences will not be detectable in MOST
observations of extrasolar planets.  However, they could be important
diagnostics in data from later space missions like Kepler and Eddington with
improved sensitivity and long-term stability.  Therefore, we have
retained these effects in our models.

Given that the orbital inclinations, radii and atmospheric structure
and compositions are unknown for most extrasolar planets, it is
important to understand how the planet scattered light curve varies
with these parameters. Here we explore parameter variation for a fixed
atmosphere model. The amplitude of the light curve is highly dependent
on the inclination. As shown in Figure 3a, the peak value can drop by
up to an order of magnitude when the inclination is changed from 90 to
50 degrees. Seager et al. (2000) have studied this effect, which will
be very important when considering possible detection of these light
curves.  The radius, as one might expect, makes a large contribution
to the amplitude of the light curve.  Over a small range of possible
planetary radii, the amplitude at all points varies proportionally to
the radius squared (Figure 3c). By comparing the effects of
inclination and radius (Figure 3b and 3c), the shape of the curve is
altered in a unique way for each parameter (given a specific
atmospheric model). From Figure 2, the inclination clearly distorts
the light curve shape while the radius simply scales the
amplitude. The overall shape and reflective properties of the planet
light curve are highly dependent on the presence of clouds in the
atmosphere (Figure 4 and also see \S2).

A change in the planet's semi-major axis would change the amplitude of
the scattered light curve by a factor of $1/a^{2}$. However, a
different semi-major axis will also change the shape of the light
curve (Figure 3b).  As the semi-major axis increases, the angular size
of the star as viewed from the planet decreases. As a result, beyond
0.07 AU distinct features of the phase function become visible because
they are not ``washed out'' by multi-directional photon trajectories
(see Figure 2).

Our model does allow for changes in the incident flux and angular size
of the star as seen from the planet in the case of a non-circular
orbit.  However, modeling the change in the planet atmosphere as a
function of its changing temperature in an eccentric orbit is much
more complicated.  Such temperature variations will also affect the
photon wavelength distribution and the level of tidal distortion.
However, only close-in extrasolar planet systems with nearly circular
orbits have been selected as MOST primary targets so the current
assumptions of zero eccentricity are valid.

\subsection{Simulations of MOST Photometry of Extrasolar Planets}

The outputs of the extrasolar planet light curve model described in \S2
and \S3 were used as the inputs to the MOST photometric simulation
program described in \S5.2. The light curve model gives the intrinsic
variability of the (star + planet) system as seen from above the
Earth's atmosphere, while the photometric simulation adds realistic
noise and variability inherent to the MOST instrument. Three different
inclinations of the planet orbit have been considered: $i =
33^{\circ}$, $i = 67^{\circ}$, and $i \sim 90^{\circ}$ (more
accurately, the maximum inclination that does not produce transits,
since transits have not been observed in $\tau$ Bo\"otis or 51 Peg).

For this paper, we present synthetic data for one of the prime
extrasolar planet targets for the MOST mission: $\tau$ Bo\"otis b. The
star $\tau$ Bo\"otis can be observed by MOST for about 50 days without
interruption (except for brief passages through the SAA; see \S4.1),
so the synthetic data set spans this time interval. The integration
time for each exposure is 24 seconds.

The reduction of the synthetic data has been fairly simple and
conservative, deliberately avoiding any calibrations that could be
influenced by our foreknowledge of the input. Mean bias values have
been subtracted from all the measurements. Exposures obviously
affected by cosmic ray strikes, and those collected during spacecraft
passage through the SAA, have been discarded. The synthetic data are
then binned to produce a net time sampling of 100 min (the orbital
period of MOST) to average out any periodic variations in stray light
and temperature due to orbital modulation.  In these simulations, we
have adopted a granulation noise amplitude and spectrum comparable to
the Sun (see \S3).

The simulated photometry for $\tau$ Bo\"otis observed at an orbital
inclination of $i = 67^{\circ}$ is presented in Figure 5, showing the
unbinned data (filled symbols) and the same data binned into groups of
100 min each (open symbols).  The modulation of the flux due to the
extrasolar planet orbit orbital period is just barely discernible by
eye in the data presented in this form.  The periodic modulation
becomes more obvious in Figure 6, where those data have been binned in
phase according to the known orbital period of $\tau$ Bo\"otis b. Also
shown in this figure are the original input models for the three
different inclinations modeled. The binned data clearly follow the
input model appropriate for this data set.

\subsection{Harmonic Structure of the Light Curves}

The detection and characterization of the planet scattered light
variation is even more obvious in Fourier space.  In
Figure~\ref{fig:MOSTamp} we show Fourier amplitude spectra of the time
series presented in Figures 5 and 6, plotted out to a frequency of
0.03 mHz.  The Nyquist frequency of the sample is 0.08 mHz, but there
are neither spectral window artifacts nor increased noise at higher
frequencies.  The inset in Figure~\ref{fig:MOSTamp} shows the spectral
window function, demonstrating that the MOST data sampling does not
introduce any serious aliasing.  These data contain intrinsic stellar
granulation noise with a 1/f frequency dependence, which is
principally evident starting at frequencies below 0.003 mHz.

The fundamental peak and characteristic harmonics in Fourier space
make even the low-amplitude periodic signals easier to recognize.
However, Figure~\ref{fig:MOSTamp} also shows that the Fourier spectrum
of the photometry is a valuable way to objectively describe the
detailed shape of the light curve.  The spectrum of the simulated MOST
data is plotted as the bold curve, while the three representative
input simulations of the planet light curves are lighter lines.  The
MOST ``data'' and the $67^{\circ}$-inclination model to which it
corresponds lie on top of one another.  Note also that the harmonic
structure of the light curves is very sensitive to the inclination.
The amplitude ratio of the first harmonic to fundamental drops
noticeably with decreasing inclination compared to higher harmonics.

To investigate this further, we generated a more complete grid of
models sampling orbital inclination $i$, for two planet radii (1.1 and
1.5 Jupiter radii), and plotted the fundamental and harmonic peak
amplitudes as a function of orbital inclination $i$
(Figure~\ref{fig:harmonic}).  This figure also quantifies our ability
to detect light variations for various inclinations and radii (for our
fixed fiducial model atmosphere).  We show in Figure 8 a very
conservative detection limit of 4.2 ppm; this is $3 \times$ the mean
noise level, corresponding to about 99.7\% confidence. We emphasize
that the detection threshold given in Figure 8 is extremely
conservative, based on the detection of signal peaks in amplitude (not
power) whose frequencies are not known a priori.  In a power spectrum,
the signal-to-noise evident in Figure 8 would be squared, but we
prefer to present amplitude spectra to err on the side of caution.
Also, we will know in advance the frequencies of the fundamental
orbital period and its harmonics, so the standard $3\sigma$ detection
limit is a severe overestimate.  Figure 8 suggests that planetary
reflected light signals should be detectable even at relatively modest
orbital inclinations.

The harmonic amplitudes have different dependences on inclination and
radius, which will be valuable in finding the correct match between
model and data.  The ``forward'' approach of adjusting the model to
fit the observations is not efficient and may lead to close but
incorrect matches.  By comparing the harmonic content of the data to
those of models from a grid of extrasolar planet parameters, we can
eliminate obvious mismatches and narrow the search to the most
promising candidate models more quickly and reliably.  This approach
is already widely used in the pulsating star community, where Fourier
decomposition of $\delta$ Scuti light curves has become a valuable
tool in identifying non-radial modes in those pulsators (e.g., Poretti
2001 and references therein).  We are exploring the diagnostic
potential of Fourier decomposition for extrasolar planet light curves,
including the underlying physics which affect the light curve shapes,
and will present this work in a subsequent paper.

\section{Discussion and Conclusions}

The first detection and measurements (even of moderate S/N) of CEGP
light curves will significantly advance our understanding of these
planets.  MOST will be the first instrument with the photometric
precision to tackle this task.  To demonstrate MOST's exciting
potential, we have run a series of simulations for a specific fiducial
atmospheric model of the planet $\tau$ Bo\"otis b (described in
\S\ref{sec-atm}).  Other atmosphere models will result in different
light curve shapes and amplitudes; however, the condensate size
distribution we have adopted is plausible for a quiescent atmosphere
(Ackerman \& Marley 2001; Cooper et al. 2003).  The parameter space of
CEGP atmospheric unknowns is so large at present (see
\S\ref{sec-uncertainties}) that a full exploration is beyond the scope
of this initial study.  MOST will soon return real data, either
measuring the albedo and the CEGP light curve shape, or setting a
meaningful upper limit.  This will greatly narrow the allowed range of
parameter space of atmospheric models.

Using our fiducial model for $\tau$ Bo\"otis b, orbiting with a period of
about 3.3 days, we have shown that MOST's conservative threshold for
detection of a light variation is about 2.5 ppm, with binned data
taken over 50 days.  This estimate includes realistic models of both
stellar granulation noise and of MOST's noise environment. Such a low
limit means we have a good chance to measure the planet light curve
even if the atmosphere differs from our fiducial model.  Furthermore,
MOST should detect the CEGPs across a relatively broad range of
orbital inclinations. The Fourier amplitude spectrum of the data will
be particularly sensitive to the signal and the detailed shape of its
light curve.

Because the actual light curve shape (and hence dominant scattering
particle type) is unknown a priori, we will need to fit many different
atmosphere models with different radii and inclinations to the real
data. Although from our simulations we can recover the fiducial input
model, including the planet radius and inclination, there is little
point specifying the accuracy of such a recovery; with real data the
goal is to detect and measure the shape of the light curve to
constrain the atmosphere model, radius, and inclination.  Although
this work indicates that the degeneracy between planet light curve,
radius, and inclination should not be severe, more work is needed to
explore this for a variety of atmosphere models.

The case of HD 209458b offers a unique opportunity to determine the
atmospheric composition because the planet's radius and inclination
are already known from fits to the transit light curve.  A measurement
of the secondary transit would give the albedo at a known phase angle
and radius.  In addition the shape of the light curve will aid us in
first determining a light curve signature to be used in detection of
light curves from planets with non-edge-on inclinations, and in
progressing towards a workable model of the atmosphere.  We are
currently working on simulations of HD 209458b.

In modeling CEGP light curves we have made several improvements and
extensions upon previous work. One significant point is that the
angular size of the star is important for planets with semi-major axes
$<$ 0.1 AU. This affects the high-angular-resolution features of the
light curve compared to using a point-source star (see
Figure~\ref{fig:lc1}). In addition, modeling the light curve with a
star of finite angular size instead of a point source causes a
reduction in amplitude of a highly backscattering-peaked light curve
by approximately 20 percent (as first noted in Seager et al. 2000).
The other new effects we investigated, tidal distortion of the planet
and stellar backheating, were found to have a negligible effect on the
planet light curve at the level of sensitivity of the MOST instrument
but may be important for subsequent space missions.

The results of this paper strongly suggest that MOST will be able to
detect the $\tau$ Bo\"otis planet light curve.  Even a null result on
this star and the other CEGP's in the MOST target list---given the
ultrahigh photometric precision attainable---would eliminate a vast
range of extrasolar planet atmosphere models with medium to high
albedos.

\begin{figure}
\plotone{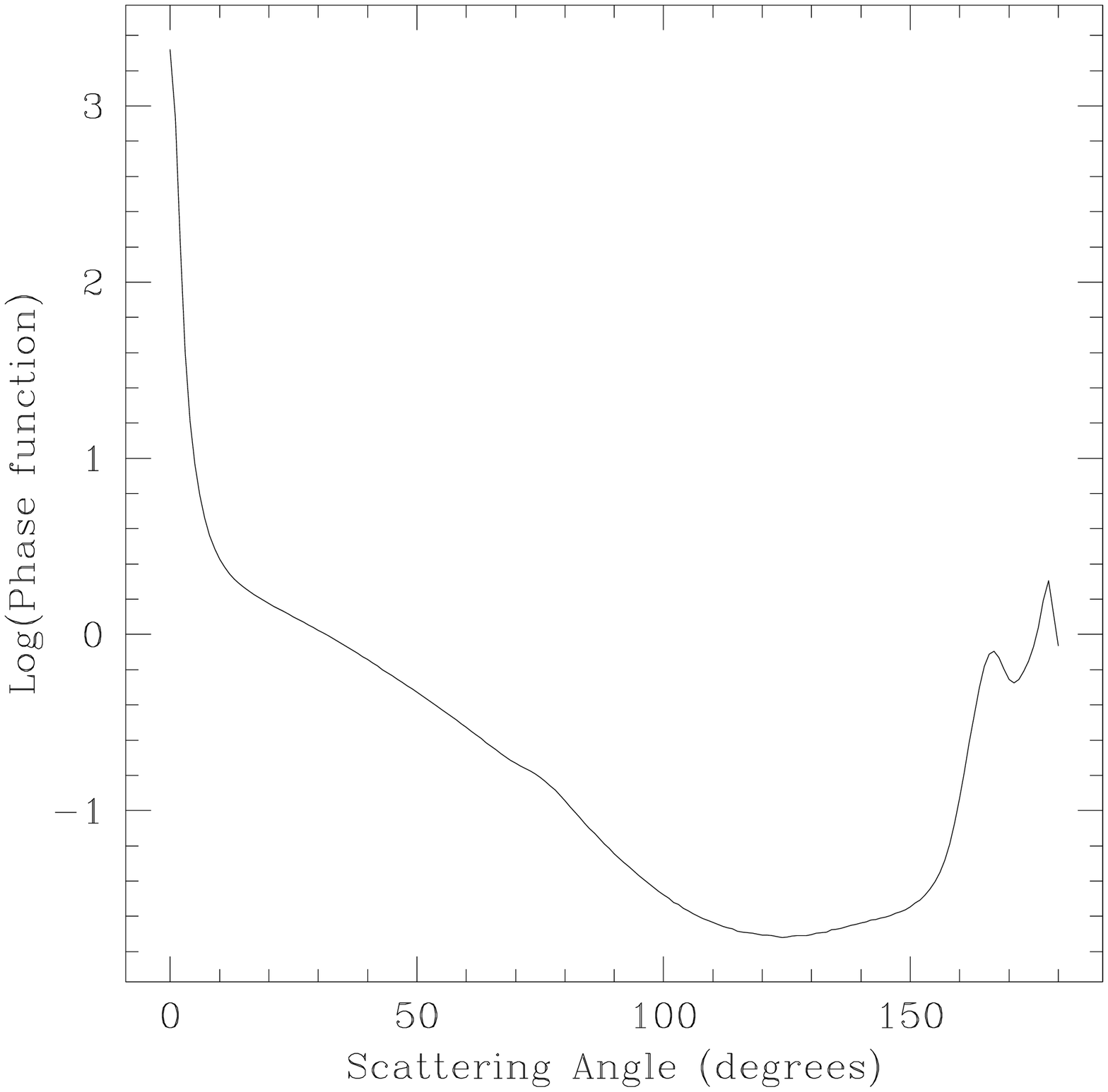}
\caption{The phase function of MgSiO$_3$ at at wavelength of 400.6
nm. We use a log-normal particle size distribution with mean radii of
5~$\mu$m and $\sigma=1.5$~$\mu$m. The phase function at other
wavelengths used are not shown.}
\label{fig:phasefn}
\end{figure}

\begin{figure}
\plotone{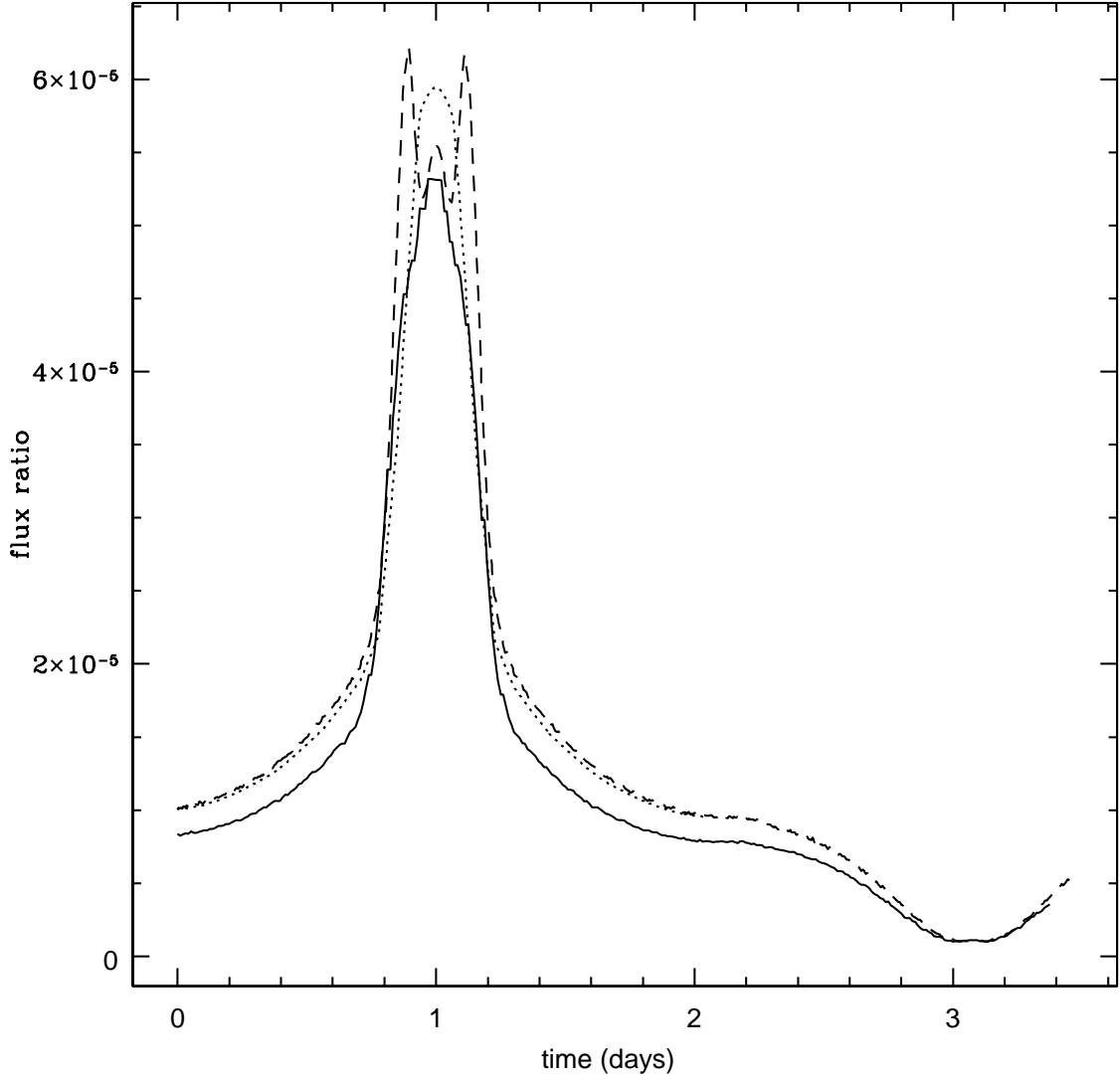}
\caption{Simulated scattered light curve of $\tau$ Bo\"otis.  The flux
ratio is given by the ratio between the reflected flux of the planet
and flux from the star. The different lines are for different parent
star assumptions: a limb darkened sphere with 0.5 degree binning
(solid line; the limb darkening is a model consistent with
measurements averaged over the visible spectrum), a point source
(dashed line) with 0.5 degree binning, and a point source with 6
degree binning in the light curve (dotted line). Note the features in
the light curve near time = 1 day when a point source (i.e.,
plane-parallel light rays) with 0.5 degree binning is used.}
\label{fig:lc1}
\end{figure}

\begin{figure}
\plotfiddle{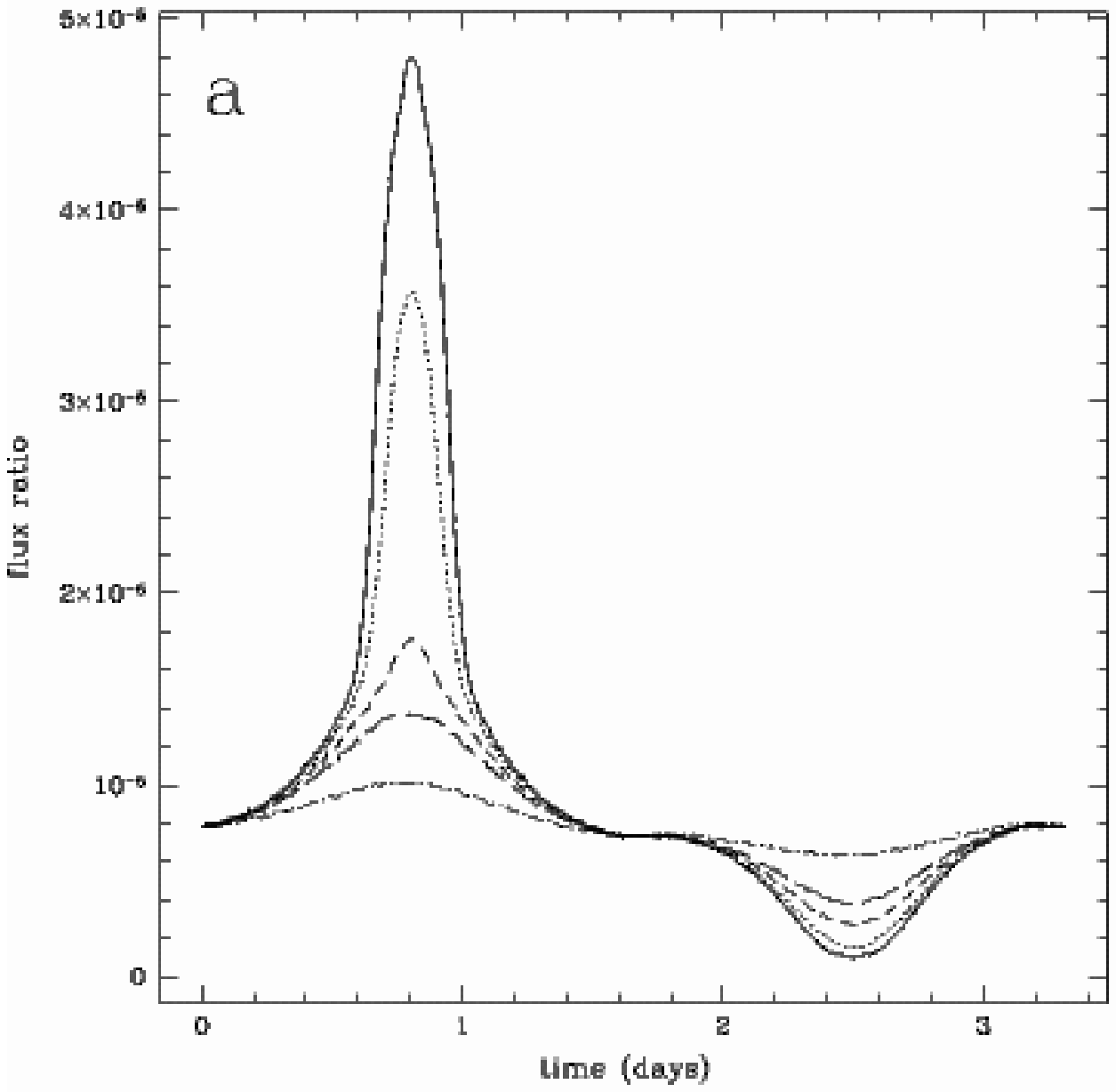}{3in}{0}{40}{45}{-240}{-80}
\plotfiddle{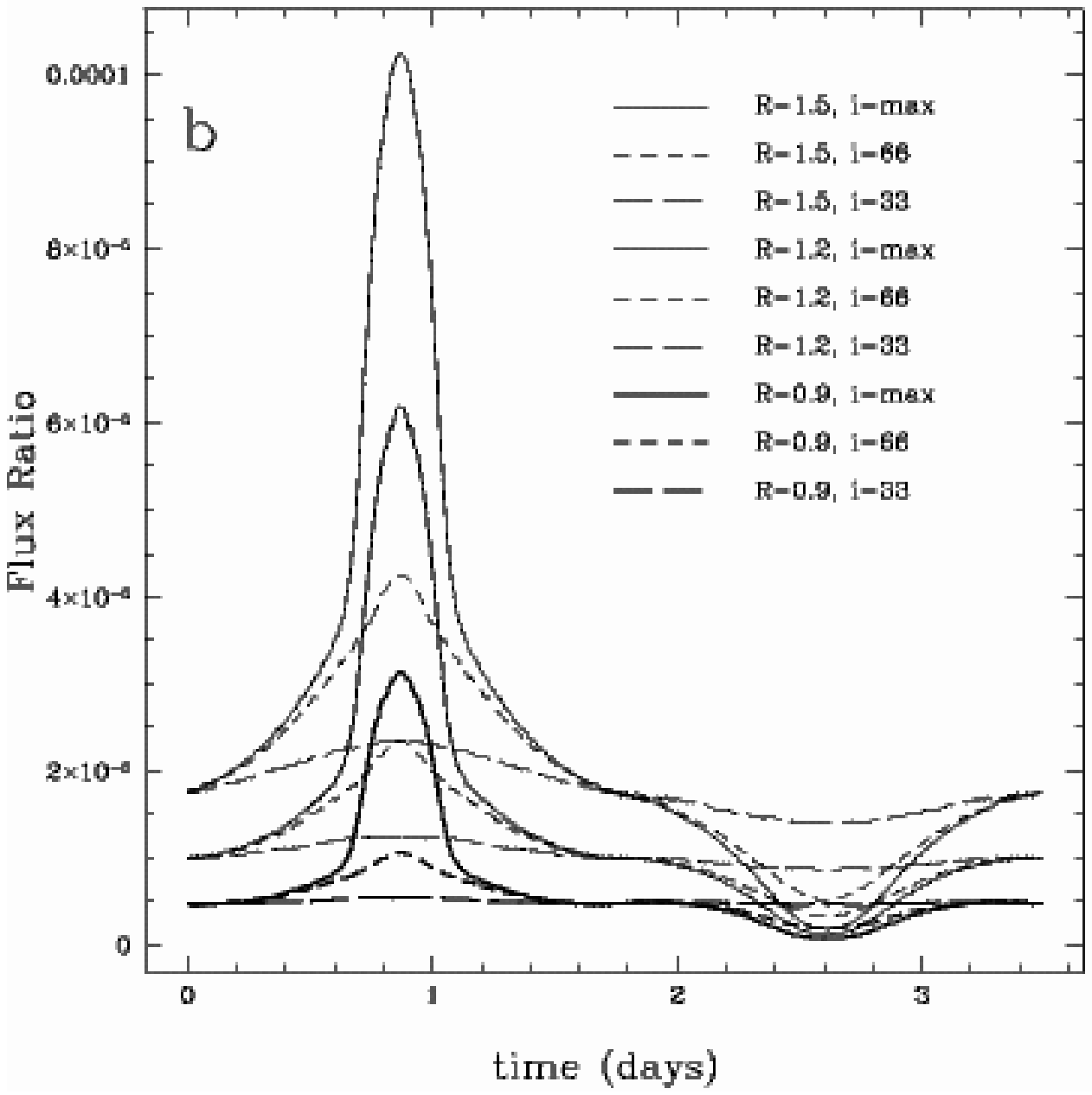}{0in}{0}{40}{46.5}{-10}{-65}
\plotfiddle{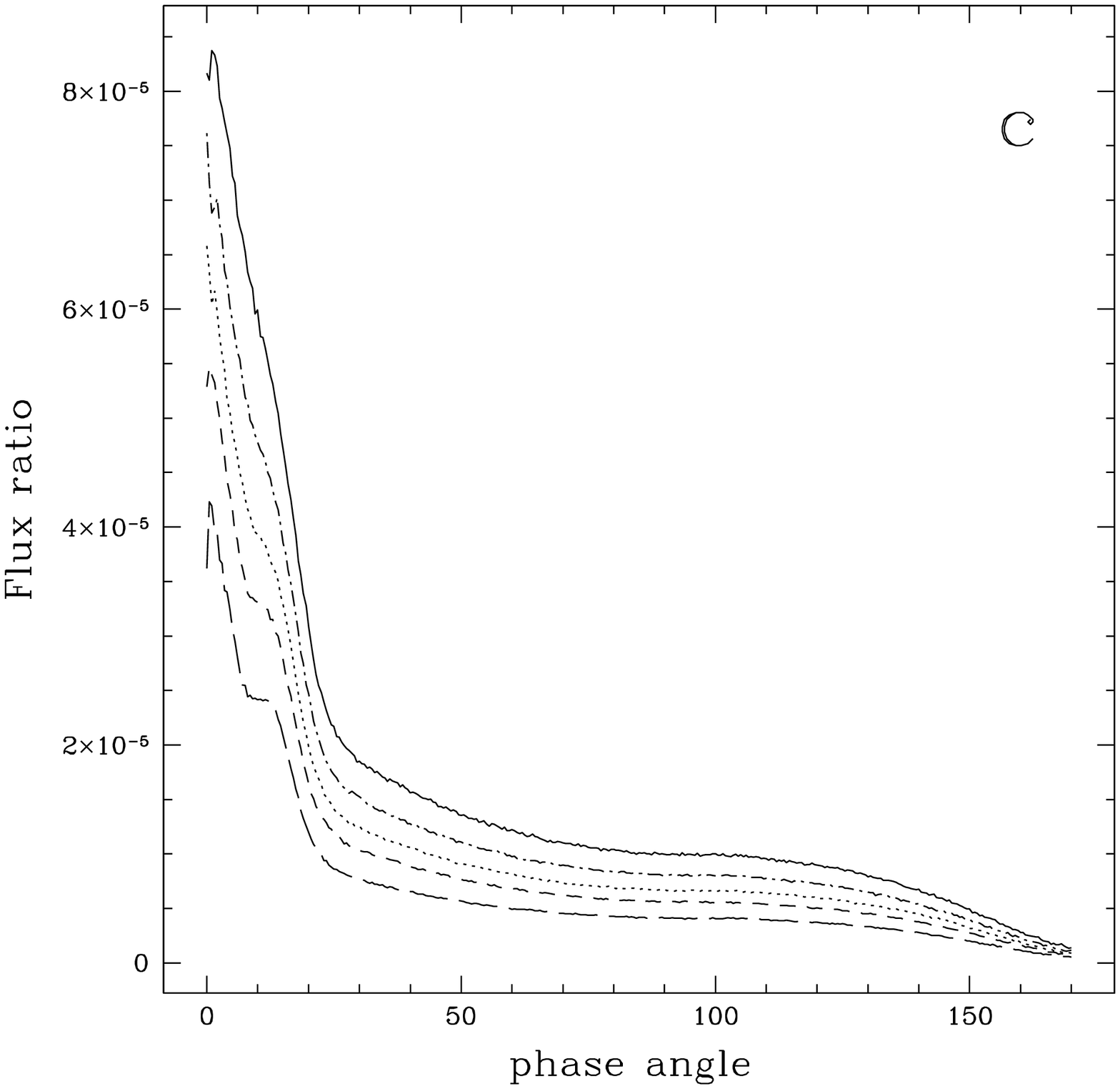}{3.5in}{0}{45}{45}{-150}{-60}
\caption{Effects of changing the inclination, semi-major axis, and
planet radius on the planet light curve for our fiducial model
atmosphere. Panel a: effects of inclination on our $\tau$ Bo\"otis model at
0.035 AU and with $R_p$ = 1.3 $R_{J}$. The different inclinations
shown are: 30 (long dash dot), 50 (short dash dot), 67 (long dash), 75
(short dash), and 80 (dot) degrees as well as the maximum before
transit (solid), 85 degrees.  Panel b: the effects of changing $R_p$
and inclination for a model planet at 0.045 AU. The planet radius is
1.5 $R_J$ (top set of curves), 1.2 $R_J$ (middle set of curves) and
0.9 $R_J$ (bottom set of curves), and inclinations of maximum before
transit (solid) and 66 degrees (dot), and 33 degrees (dashed). Note
that the planet atmosphere model was not changed to account for the
different $R_p$. Panel c: effects of different semi-major axes on our
$\tau$ Bo\"otis model of 1.3 $R_{J}$ at maximum inclination before
transit. The semi-major axes are 0.045 AU (solid), 0.05 AU (dot), 0.06
AU (short dash) and 0.07 AU (long dash). Note that the planet atmosphere
model was not changed to account for the different semi-major axes and
parent star irradiation.}
\label{fig:lcmulti}
\end{figure}

\begin{figure}
\plotone{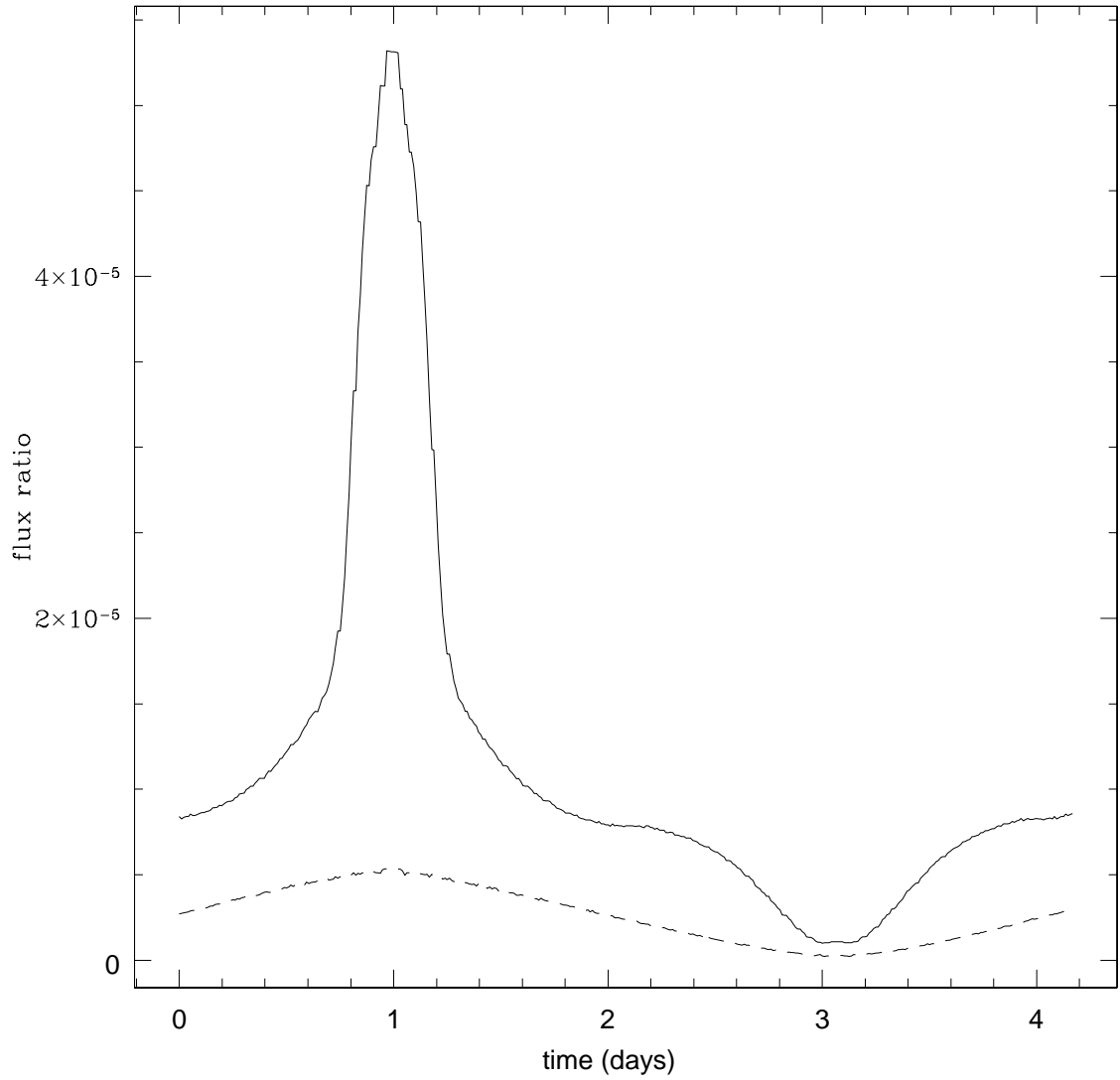}
\caption{A model planet at 0.05 AU and $R_p$ = 1.3 $R_{J}$ is shown with
(solid) and without (dash) a cloud layer.}
\label{fig:lcr}
\end{figure}

\begin{figure}
\plotfiddle{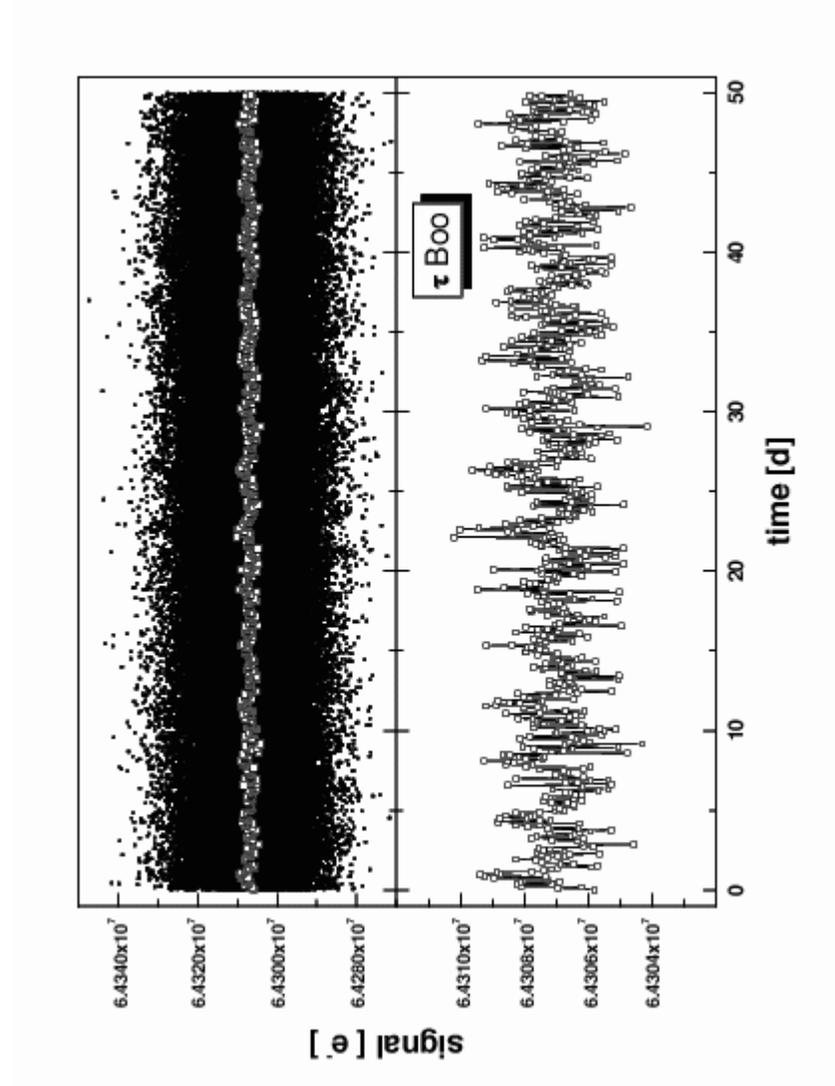}{7in}{0}{75}{75}{-200}{-30}
\caption{Simulated MOST photometry of $\tau$ Bo\"otis, 67 deg inclination
model, time base 50 days.  Upper panel: small black squares, signal in
[e-] for 25 seconds integrations (data collected in the SAA or
affected by cosmic rays have been rejected). Open squares, mean signal
[e-] data binned over the spacecraft orbit period of 100 minutes.
Lower panel is an expanded view of the binned data.}
\label{fig:MOSTsim1}
\end{figure}

\begin{figure}
\plotfiddle{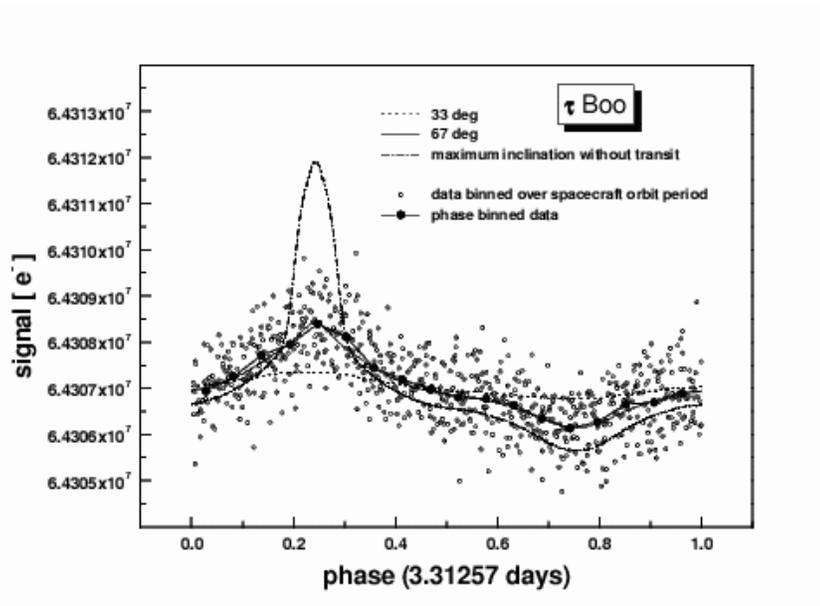}{4in}{270}{65}{65}{-250}{370}
\caption{Signals of $\tau$ Bo\"otis model (67$^{\circ}$ inclination)
photometry versus phase of the planet's orbital period.  The open
circles represent the binned data, the connected black circles are the
mean photometric signals for each (0.05) phase interval. The light
curve for the 3 models are shown as well: the dotted line is for
33$^{\circ}$, the solid line for 67$^{\circ}$ deg and the dash dot
line for the maximum inclination before transit.}
\end{figure}

\begin{figure}
\plotone{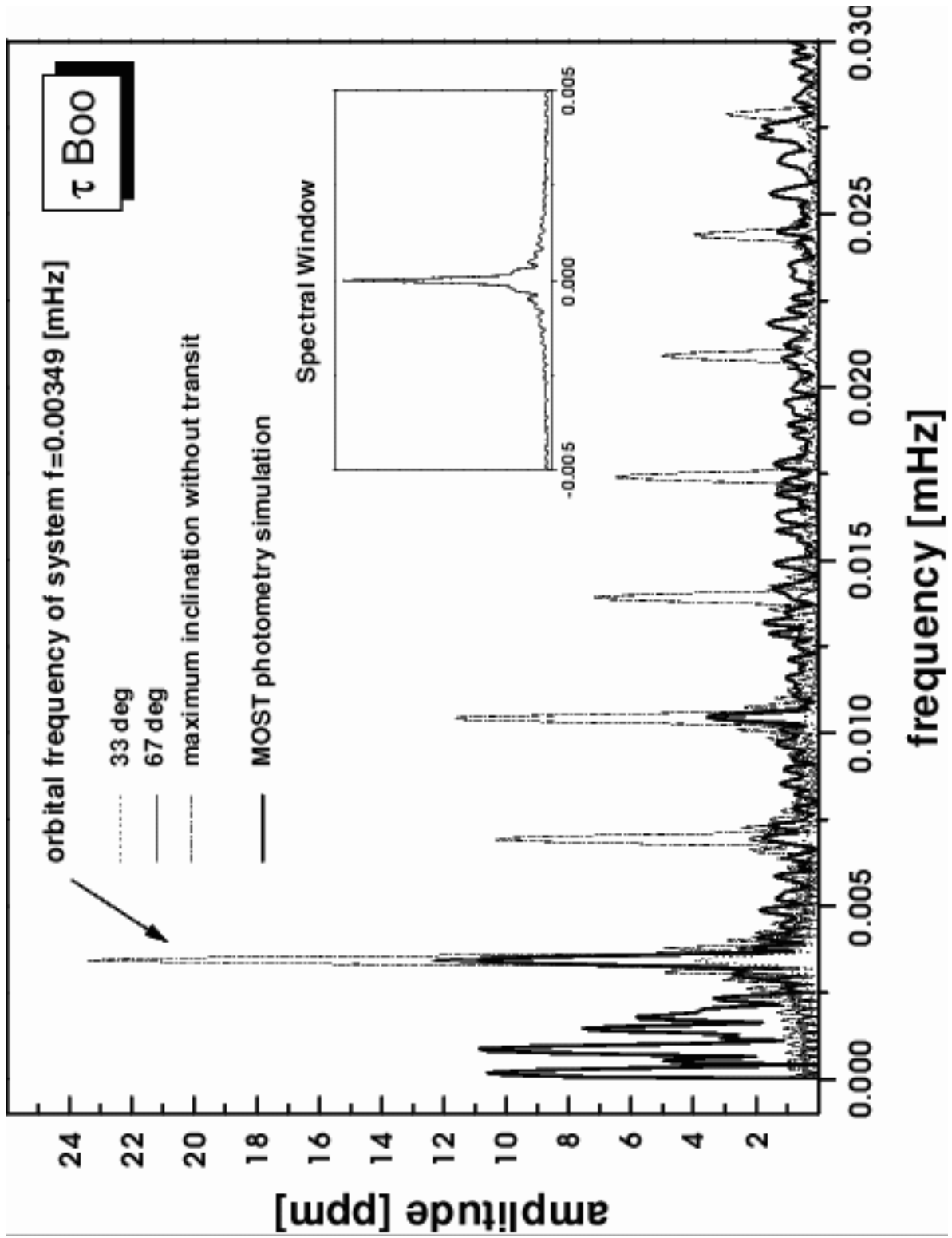}
\caption{Amplitude spectra a) of a $\tau$ Bo\"otis model (67$^{\circ}$
inclination), b) photon, instrumental and granulation noise, c)
combined. In addition the spectral window is shown (upper panel).  The
Fourier analysis of the noise has been applied to the binned data with
a time base of 50 days.}
\label{fig:MOSTamp}
\end{figure}

\begin{figure}
\plotfiddle{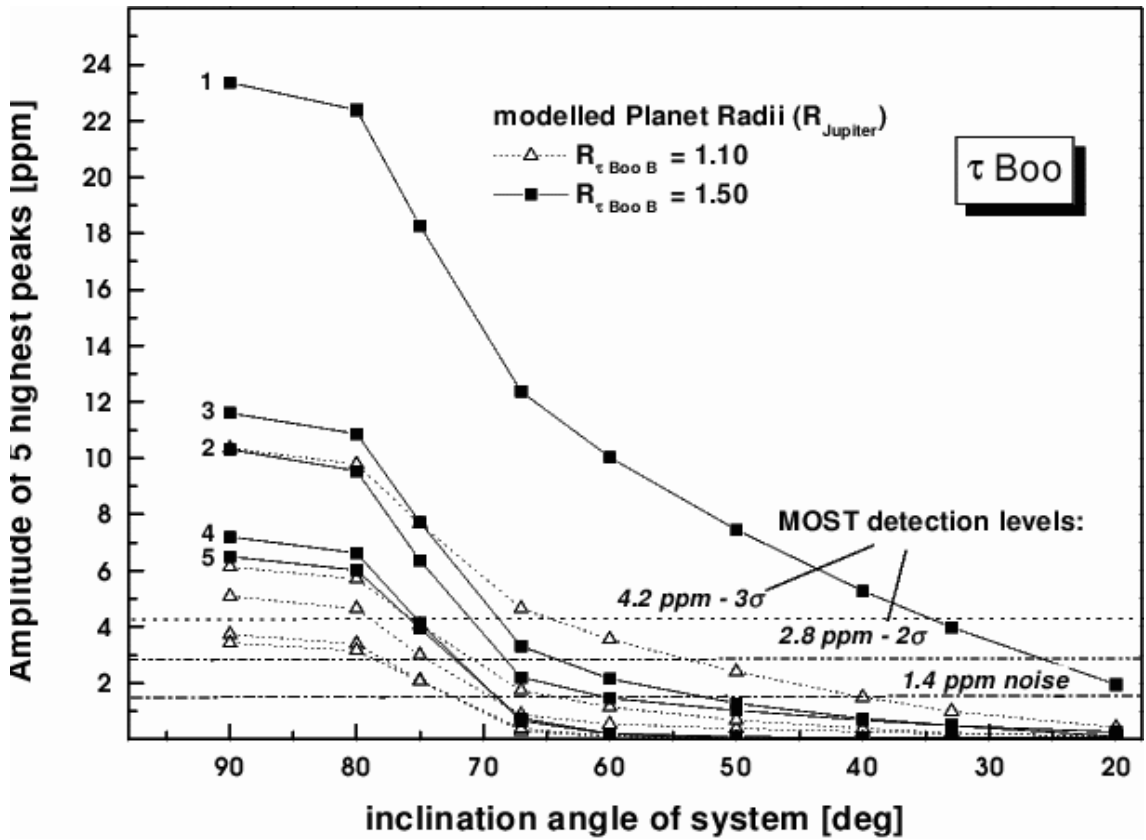}{7in}{270}{70}{70}{-280}{450}
\caption{Amplitude of the five highest Fourier peaks
as a function of orbital inclination (corresponding
to the fundamental and the four lowest harmonics relative to the
fundamental) for the $\tau$ Bo\"otis amplitude spectrum shown in
Figure 7.  The horizontal lines (from bottom to top) show the mean
noise level, and the 2$\sigma$ and 3$\sigma$ detection limits.  (Note
that in the power spectrum, as opposed to this amplitude spectrum, the
signal-to-noise will be squared.)}
\label{fig:harmonic}
\end{figure}

\end{document}